\documentclass[prb,superscriptaddress,twocolumn]{revtex4}
\usepackage[utf8]{inputenc}
\usepackage{amsmath}
\usepackage{graphicx}
\usepackage{multirow}
\usepackage{color}
\usepackage{hyperref}
\begin{document}
\definecolor{orange}{rgb}{1, 0.5, 0.0}
\title{Optical-like interference control polar-phonons dispersions in 2D materials}

\author{Benoit Van Troeye}
\affiliation{Imec, Leuven, Belgium}
\author{Geoffrey Pourtois}
\affiliation{Imec, Leuven, Belgium}
\begin{abstract}
   It has been argued that the polar phonon modes of two-dimensional (2D) materials should show a linear dependence close to the Brillouin zone center due to the reduced problem dimensionality compared to a three-dimensional crystal, leading to a vanishing LO-TO splitting. Revisiting this question using a mirror-charge framework derived from classical electrostatics, we show that it is not always true: with perfectly-reflective dielectric interfaces, it is possible to recover a LO-TO splitting, or an equivalent phenomenon for the out-of-plane modes. Effectively, the dispersion of polar phonon modes is governed by optical-like constructive and destructive interferences of the source potential with its reflection at the dielectric interfaces. This work highlights the critical role of dielectric boundary conditions to understand phonon-related properties in 2D materials.
\end{abstract}
\maketitle

Accurately predicting carrier mobilities in two-dimensional (2D) materials requires a proper treatment of the long-range electrostatics to their Interatomic Force Constants (IFCs) and of the associated electron-phonon potential using first-principles methods~\cite{Ponce2023,Ponce2023b}. Theory shows that the dispersion of polar 2D phonon modes becomes linear in the immediate vicinity of the Brillouin zone center~\cite{Sohier2017,Sohier2017b,Royo2021}, in stark contrast to the three-dimensional case, where a clear splitting between the Longitudinal Optical (LO) and  Transverse Optical (TO) modes is observed~\cite{Gonze1997}. This deviation has been attributed to the dimensionality reduction from 3D to 2D~\cite{Sohier2017,Royo2021}. Moreover, theoretical work predicts that the phonon dispersion is strongly influenced by its surrounding dielectric environment~\cite{Sohier2017}, with the polar-mode dispersion being fully quenched when the 2D material is placed in contact with a metal. This effect has recently been confirmed experimentally~\cite{Li2024}. 

In this work, we revisit these theoretical predictions using the concept of mirror charges derived from classical electrostatics, providing a clear physical explanation for the above observations and separating the interface-related contributions from the intrinsic ones. When a polarization is induced in a 2D material by an atomic displacement, part of the resulting electric field is reflected at the surrounding dielectric interfaces, generating optical-like constructive and destructive interferences with the source polarization. In general, the losses due to the multiple reflection at small phonon wavevector are too high to sustain a proper LO-TO splitting, and lead to the linear dispersion. 
We propose a scheme to correct long-range contributions to the Interatomic Force Constants (IFCs) obtained from ab initio calculations including the contribution from to the dielectric environment. The applicability of such a scheme is demonstrated for several representative 2D materials, including monolayer hexagonal boron nitride (hBN), phosphorene, MoS$_2$ and HfS$_2$.

\begin{figure}[h]
\includegraphics[width=0.37\textwidth]{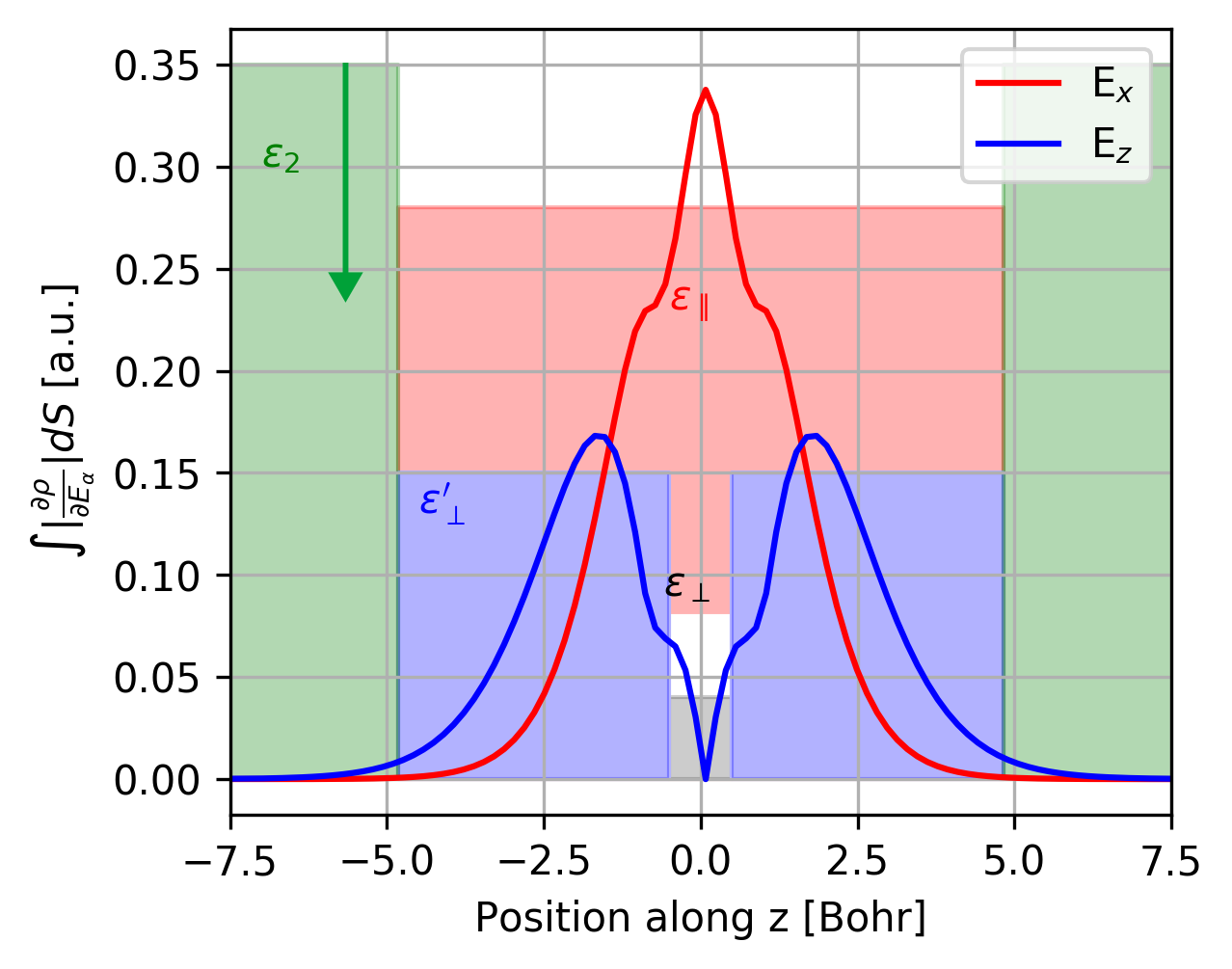}
\caption{Normalized and absolute change of electronic density along the out-of-plane direction in hBN monolayer with an external electric field (E) from ab initio calculations (integrated in-plane for surface S). While most of the screening occurs close to atomic layer for an in-plane electric field, the out-of-plane response is in contrast located further away from it. This observation motivates to use a more complex model for the dielectric response of the 2D material along the z direction, with two different dielectric constants $\epsilon_{\perp}$ and $\epsilon'_{\perp}$ (blue and white areas, respectively), while the in-plane response can be approximated with a single dielectric constant $\epsilon_\parallel$ (read area). The dielectric constant of the environment is then tunable in the model.}
\label{fig:changedensity}
\end{figure}

In 3D crystal, the standard approach relies on treating separately the long-range IFCs with classical electrostatics and the Ewald summation~\cite{Gonze1997}, which leads to the typical LO-TO splitting. In the 2D case, an additional challenge arises from the fact that the dielectric response of the 2D material does not extend indefinitely along the out-of-plane direction, and thus the Poisson equation cannot so straightforwardly solved. M. Royo and M. Stengel provided a rigorous formulation of an effective dielectric model based on the irreducible response and open boundary conditions~\cite{Royo2021}, in which the out-of-plane structure of the dielectric response is incorporated into effective parameters. Here, in closer connection with the approach of Sohier \textit{et al.}~\cite{Sohier2017,Sohier2017b}, we instead retain explicitly the boundary conditions and the z-dependence of the dielectric response as a first step, and work in terms of the reducible Coulomb potential.
Taking the out-of-plane axis as z, and assuming that the dielectric tensors have no mixed in-plane/out-of-plane components ($\epsilon_{xz}=\epsilon_{yz}=0$), the electrostatic potential satisfies in mixed periodic in-plane and non-periodic out-of-plane coordinates:
\begin{equation}
\mathbf{G}_{\parallel}.\underline{\underline{\epsilon}}_\parallel\mathbf{G}_{\parallel} \tilde{V}(z)-\frac{\partial}{\partial z}\left[\epsilon_\perp(z)\frac{\partial \tilde{V}}{\partial z}\right]=4\pi\tilde{\rho}(z)
\end{equation}
where $\mathbf{G}_{\parallel}$ are the in-plane reciprocal lattice vectors, $\underline{\underline{\epsilon}}_\parallel$ and the in-plane components of the dielectric tensor, $\tilde{V}(z)$ and $\tilde{\rho}(z)$ are the Fourier components of the potential and of the charge density. When examining the microscopic response with the first-order change of the electronic density with an external electric field for hBN (see Fig.~\ref{fig:changedensity}), approximating the dielectric response of the 2D material with a single in-plane dielectric constant appears appropriate to reproduce the in-plane response, while the out-of-plane screening emerges away from the atomic z-positions. In this later case, the dielectric response is refined by considering two consecutive dielectric slabs, as schematically shown in Fig.~\ref{fig:changedensity}. This piece-wise decomposition is important to solve the Poisson's equation analytically.

The change of charge density following an atomic displacement is extended in the long-wavelength limit, following Refs.~\onlinecite{Royo2019,Royo2022}, then converted in the mixed coordinates:
{\footnotesize
\begin{multline}
\tilde{\rho}(z)=-\left[j\mathbf{G}.\mathbf{Z}^*_{\kappa\alpha,\parallel}+Z^*_{\kappa\alpha z}\frac{\partial}{\partial z}+\frac{1}{2}\mathbf{G}.\underline{\underline{Q}}   \,_\mathbf{\kappa\alpha,\parallel\parallel}\mathbf{q} -j\mathbf{G}.\mathbf{Q}_{\kappa\alpha, \parallel z}\frac{\partial}{\partial z}\right.\\
\left. -\frac{1}{2}\mathbf{Q}_{\kappa\alpha,zz}\frac{\partial^2}{\partial z^2}+\cdots\right]\delta(z-R_{\kappa z}) e^{j\mathbf{G}.\mathbf{R}^\mathbf{b}_{\kappa \parallel}} u^\mathbf{b}_{m\mathbf{q}}(\kappa\alpha)
\end{multline} }
where $Z^*_{\kappa\alpha.\beta}$ are the Born effective charges ($\parallel$ indicates the directions are only considered in-plane), $Q_{\kappa\alpha,\beta\gamma}$ are the dynamical quadrupole moments~\cite{Royo2019,Royo2020b,Royo2022}, $\alpha,\beta,\gamma$ are directions, $\mathbf{R}^b_{\kappa \parallel}$ is the in-plane position of atom $\kappa$ in cell $\mathbf{b}$, and $u^\mathbf{b}_{m\mathbf{q}}$ is the phonon-mode displacement of atom $\mathbf{\kappa}$ along the phonon mode $m$ at wavevector $\mathbf{q}$. Because phonon wavevectors are typically much larger than those of photons, phonon-photon coupling can be safely neglected and the problem may be treated entirely within an electrostatic framework. 

\begin{figure}[h]
\includegraphics[width=0.48\textwidth]{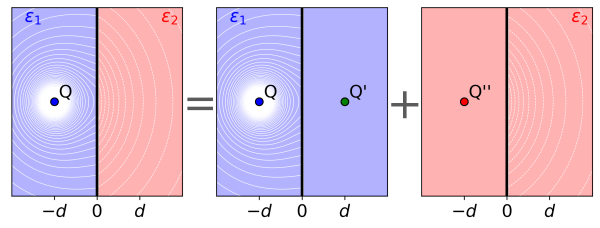}
\caption{Schematic illustration of the mirror-charge concept at the interface between two dielectrics. White lines indicate the electrostatic-potential isolines. Since the Poisson equation must be solved piecewise, the potential is defined separately in each region. For z$<$0, the potential corresponds to the superposition of two charges Q and Q', embedded in dielectric $\epsilon_1$ and located symmetrically with respect to the interface; these represent the source and its reflected image. For $z>0$, the potential is generated by a single effective charge located at $z=-d$, which accounts for the transmitted contribution across the interface.  }
\label{fig:mirrorcharge}
\end{figure}

The problem becomes closely analogous to the classical mirror-charge construction described in standard physics textbooks~\cite{Griffiths2013}, a concept we briefly summarize here. We consider an interface separating two semi-infinite  dielectrics with isotropic permittivities $\epsilon_1$ and $\epsilon_2$ and a point charge located in first medium at $z=-d$. To impose continuity of the tangential electric field and of the normal displacement field, one needs to introduce additional fictitious charges (known as mirror charges) in the system, positioned at the source charge and at the symmetric point across the dielectric interface (see Fig.~\ref{fig:mirrorcharge}). Physically, the former mirror charge leads to the transmitted potential outside of the first dielectric, while the charge placed on the opposite side of the interface generates the reflected contribution at the boundary due to the dielectric mismatch. Consequently, and in contrast to the case of a homogeneous medium, the original charge interacts with its own reflected field at the interface. Depending on the relative permittivities $\epsilon_1$ and $\epsilon_2$, it leads to optical-like constructive and destructive interferences.

We then refine this mirror-charge concept to our original statement: describing the long-range electrostatic contributions to the IFCs in 2D materials. To achieve this, we extend the classical mirror-charge problem to account for several key features relevant to the problem: i) the intrinsic anisotropy of the 2D dielectric tensor, ii) the treatment of electrostatic divergence using the Ewald summation~\cite{Martin2004}, iii)  the presence of a dielectric slab embedded in a homogeneous dielectric environment, iv) the connection to Born effective charges and dynamical quadrupoles, and v) consecutive dielectric slabs. The corresponding mathematical developments are provided in the Supplementary Materials (S.M.).

After performing the corresponding derivations, we obtain an analog expression for the IFCs in 2D for the reciprocal summation of Ewald summation that incorporates the contributions (here, restricted to dipole-dipole for brevity):

\begin{widetext}
\small{\begin{equation}
\begin{array}{l l}
\tilde{\Phi}^{\text{DD,rec}}_{\kappa\alpha\kappa'\beta}(\mathbf{q})=\frac{2\pi}{S\epsilon_\perp}&\sum_{\mathbf{K}=\mathbf{G}_\parallel+\mathbf{q}}  \left\{\frac{\mathbf{K}.\mathbf{Z}^*_{\kappa\alpha}\mathbf{K}.\mathbf{Z}^*_{\kappa'\beta}}{\eta}\left[f_\text{ewald}(R_{\kappa' z}-R_{\kappa z})+\frac{2re^{-\eta d}}{1-r^2 e^{-2\eta d}}\left(\cosh(\eta(R_{\kappa z}+R_{\kappa' z}))+re^{-\eta d}\cosh{(\eta(R_{\kappa' z}-R_{\kappa z}))}\right)\right]\right.\\
& +\eta Z^*_{\kappa\alpha,z}Z^*_{\kappa'\beta,z}\left[-f_\text{ewald}(R_{\kappa' z}-R_{\kappa z})+\frac{2re^{-\eta d}}{1-r^2 e^{-2\eta d}}\left(\cosh(\eta(R_{\kappa z}+R_{\kappa' z}))-re^{-\eta d}\cosh{(\eta(R_{\kappa' z}-R_{\kappa z}))}\right)\right]\\
&-j(\mathbf{K}.\mathbf{Z}^*_{\kappa\alpha}Z^*_{\kappa'\beta,z}+\mathbf{K}.\mathbf{Z}^*_{\kappa'\beta}Z^*_{\kappa\alpha,z})\left[\frac{1}{\eta}f'_\text{ewald}(R_{\kappa' z}-R_{\kappa  z})+\frac{2r^2e^{-2\eta d}}{1-r^2 e^{-2\eta d}}\sinh{(\eta(R_{\kappa' z}-R_{\kappa z}))}\right]\\
&\left.-j(\mathbf{K}.\mathbf{Z}^*_{\kappa\alpha}Z^*_{\kappa'\beta,z}-\mathbf{K}.\mathbf{Z}^*_{\kappa'\beta}Z^*_{\kappa\alpha,z})\frac{2re^{-\eta d}}{1-r^2 e^{-2\eta d}}\sinh{(\eta(R_{\kappa  z}+R_{\kappa' z}))}\right\} e^{j\mathbf{K}.(\mathbf{R}_{\kappa \parallel}-\mathbf{R}_{\kappa' \parallel})} \\ \label{eq:ifc_2D}
\end{array}
\end{equation}}\end{widetext}
with $S$ is the unit cell surface and
\begin{equation}
\eta = \sqrt{\frac{\mathbf{K}_.\underline{\underline{\epsilon}}_{\parallel} \mathbf{K}}{\epsilon_\perp}}.
\end{equation}
The reflection coefficient 
\begin{equation}
r=\frac{\epsilon_{\hat{\mathbf{K}}}-\epsilon_2}{\epsilon_{\hat{\mathbf{K}}}+\epsilon_2},
\end{equation}
is related to the dielectric constant of the environment $\epsilon_2$ and the effective dielectric constant for the 2D material along the $\hat{\mathbf{K}}_\parallel$ direction
\begin{equation}
\epsilon_{\hat{\mathbf{K}}}=\sqrt{\epsilon_\perp}\frac{\sqrt{\mathbf{K}_.\underline{\underline{\epsilon}}_{\parallel} \mathbf{K}}}{K}.
\end{equation}
Finally, $f_\text{ewald}(z)$ and its spatial derivative $f'_\text{ewald}(z)$ denote the separation functions appropriate for the bi-dimensional case considering Gaussian charges in Ewald summation: 
\begin{multline}
    f_{\text{ewald}}(z)= \frac{1}{2}e^{-\eta z}\text{erfc}\left(\frac{ \eta \sqrt{\epsilon_\perp}\Lambda}{\sqrt{2}}-\frac{z}{\Lambda\sqrt{2\epsilon_\perp}}\right)  \\
    + \frac{1}{2}e^{\eta z}\text{erfc}\left(\frac{ \eta \sqrt{\epsilon_\perp}\Lambda}{\sqrt{2}}+\frac{z}{\Lambda\sqrt{2\epsilon_\perp}} \right).\\
\end{multline}

Here, $\Lambda$ corresponds to the Gaussian broadening intrinsic to the Ewald summation technique, $\text{erfc}$ is the complementary error function. 
The real-space part of the Ewald summation in 2D is identical to that of the 3D case (see Eqs.~73-74 of Ref.~\onlinecite{Gonze1997} or Eq.~S37), except that the summation is restricted to in-plane lattice vectors. As in 3D, the parameter $\Lambda$ is selected in such a way that the real-space summation becomes negligibly small. After evaluating the 2D Ewald summation, translational invariance can be enforced by applying:
\begin{equation}
\tilde{\Phi}^{\text{new}}_{\kappa\alpha\kappa'\beta}(\mathbf{q})=\tilde{\Phi}^{\text{old}}_{\kappa\alpha\kappa'\beta}(\mathbf{q})-\delta_{\kappa\kappa'}\sum_{\kappa''}\tilde{\Phi}^{\text{old}}_{\kappa\alpha\kappa''\beta}(\mathbf{0}).
\end{equation}
Note that for the out-of-plane response, a more advanced model was constructed with two consecutive dielectric slabs (see Sec.~SV) but it doesn't change fundamentally the physics of the problem. We also include dynamical quadrupoles (see Eq.~S41), but similarly to Ref.~\onlinecite{Ponce2023}, their effects are found to be small for phonon dispersions.
 
With Eq.~\ref{eq:ifc_2D}, we recover the linear behavior of the IFCs near the Brillouin-zone center in the absence of dielectric mismatch ($r=0$), consistent with previous theoretical studies~\cite{Sohier2016,Sohier2017,Royo2021}. This linear dispersion is obtained for both for the LO and ZO modes in the case of a flat 2D material. Introducing a non-zero dielectric mismatch modifies this behavior for the LO mode with the slope of the linear dispersion being enhanced for positive  dielectric mismatch ($r>0$) or suppressed for negative mismatch ($r<0$). The ZO modes exhibit the opposite behavior, with a downward shift of the ZO mode at the zone-center when interfaced with a metal.

When vacuum is used as the surrounding dielectric, the reflection coefficient is large for most materials but still smaller than unity. As a result, part of the field is transmitted outside of the 2D material layer, preventing strong constructive interferences at long phonon wavelengths. Losses introduced by repeated reflections at the interfaces suppress the buildup of coherent interactions over long distances, explaining the non-existence of the LO-TO splitting in general for 2D materials. However, such a finite LO-TO splitting can be restored by appropriately engineering the dielectric response of the surrounding medium. In particular, when $\epsilon_{2}\rightarrow0$, we obtain a divergence of the electrostatic potential for long-wavelength phonons analogous to the 3D case and thus a LO-TO splitting. This suggests that by tuning the plasmon frequency of nearby metals, for example graphene under electrostatics gating~\cite{Ju2011,Grigorenko2012,Low2014}, a LO-TO splitting could, in principle, be observed experimentally in 2D materials.  

Building on these theoretical developments, we have implemented the corresponding expressions in the ABINIT software package~\cite{Abinit2005,Abinit2025}. The short-range component of the IFCs is first extracted from the full ab initio IFCs by subtracting the long-range contribution of Eq.~\ref{eq:ifc_2D} (or Eq.~S47 for refined model), evaluated using vacuum as the dielectric environment:
\begin{equation}
\tilde{\Phi}^{\text{sr}}_{\kappa\alpha,\kappa'\beta}(\mathbf{q}_c)=\tilde{\Phi}_{\kappa\alpha,\kappa'\beta}(\mathbf{q}_c)-\tilde{\Phi}^{\text{lr}}_{\kappa\alpha,\kappa'\beta}(\mathbf{q}_c,\epsilon_2=1)
\end{equation}
where $\mathbf{q}_{c}$ phonon wavevector belongs to the coarse phonon wavevector grid. After the Fourier interpolation of the short-range IFCs, the analytical long-range electrostatic term is added back in, now allowing for an arbitrary dielectric environment:
\begin{equation}
\tilde{\Phi}_{\kappa\alpha,\kappa'\beta}(\mathbf{q}_d)=\tilde{\Phi}^{\text{sr}}_{\kappa\alpha,\kappa'\beta}(\mathbf{q}_d)+\tilde{\Phi}^{\text{lr}}_{\kappa\alpha,\kappa'\beta}(\mathbf{q}_d,\epsilon_2),
\end{equation}
with $\mathbf{q}_d$ belonging to the dense (interpolated) phonon wavevector mesh. The center of the dielectric slab is defined as the average atomic coordinate along the z direction and the dielectric thickness is extracted directly from the out-of-plane dielectric response.

To validate our implementation, we applied it to a representative set of 2D materials: hBN, phosphorene, MoS$_2$ and HfS$_2$. 

\begin{figure*}[htp]
\includegraphics[width=0.9\textwidth]{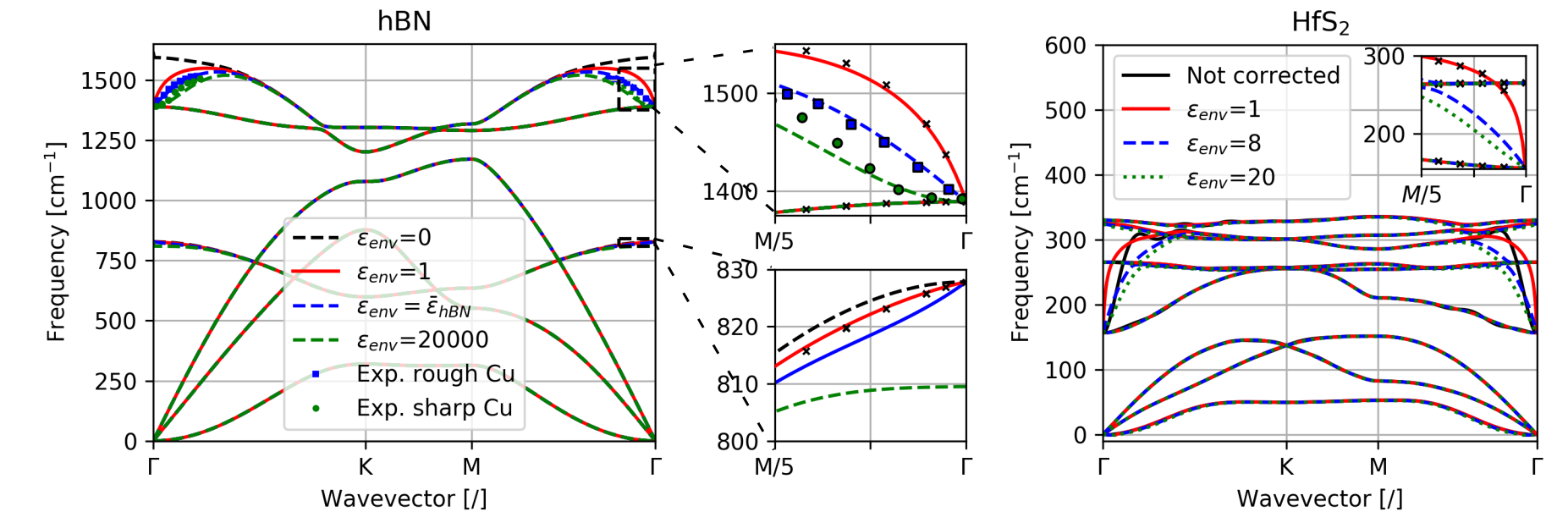}
\caption{Phonon band structures for hBN and HfS$_2$ as function of the dielectric environment. Results are shown with the long-range electrostatic treatment introduced in this work, which explicitly include the dependence on the dielectric constant of the environment (solid red, dashed blue and dotted green lines). Black crosses correspond to exactly computed points with Quantum Espresso~\cite{Sohier2016}. For hBN, we also include the experimental data from Ref.~\onlinecite{Li2024}, corresponding to the samples transferred onto a rough (down-pointing blue triangles) or sharp (up-pointing green triangles) copper substrate. In this case, the dashed black and full blue lines correspond to environments with $\epsilon_2=0$ or $\epsilon_2$ given by the geometric mean of the in-plane and out-of-plane of hBN, i.e., $\epsilon_2 =\sqrt{\epsilon_{\parallel}\epsilon_{\perp}}$ (no reflection at interfaces), respectively. For HfS$_2$, the solid black line corresponds to the case without a separated treatment of the long-range electrostatics.\label{fig:phonon_comp}}
\end{figure*}

Computational details are provided in the supplementary materials. 
The influence of the dielectric environment is particularly pronounced in HfS$_2$ and hBN, where the frequencies of the polar modes shift by several tens to hundreds of cm$^{-1}$ (see Fig.~\ref{fig:phonon_comp}), and weaker in MoS$_2$ and phosphorene (shown in the S.M. consequently). 
For hBN, we highlight a shift of approximately 20~cm$^{-1}$ in the ZO-mode frequency when placing the material in a metallic environment. We also compare our calculations to the experimental data obtained for monolayer hBN transferred onto either a rough copper foil or a flat single-crystal copper substrate~\cite{Li2024}. The latter is expected to form a clean and well-defined dielectric interface with the 2D layer. For the rough copper surface, the experimental data do not show constructive or destructive interference effects, closely matching the case of zero dielectric mismatch in our calculations. In this situation, no coherent reflection occurs at the dielectric interfaces, in agreement with our simulated case.

In conclusion, polar phonons in 2D materials should not be viewed as an intrinsic property of the isolated layer but rather as a function on the dielectric environment. 

In some specific configurations, a LO-TO splitting can even be restored. These findings have important implications for carrier-transport modeling. Even in vacuum, long-range electrostatic contributions to the IFCs have been shown to substantially influence charge carrier mobilities in 2D materials (e.g., $\sim 50 \%$ for the hole mobilities of MoS$_2$)~\cite{Ponce2023,Ponce2023b}. Our results indicate that such conclusions must now be revisited to include the explicit influence of the dielectric environment. 
The framework developed here provides a route —once adapted to electron‑phonon coupling— to systematically reassess mobility in technologically relevant dielectric configurations.

\begin{acknowledgments}
The authors acknowledge the Imec Industrial Affiliation Program (IIAP) for funding. We also thank Miquel Royo and Massimiliano Stengel for making the the Coulomb truncation implementation in ABINIT available to us for this publication.
\end{acknowledgments}

\bibliography{Biblio.bib}

@book{Griffiths2013,
	title = {Introduction to {Electrodynamics}},
	isbn = {978-0-321-85656-2},
	url = {https://books.google.be/books?id=AZx_zwEACAAJ},
	publisher = {Pearson},
	author = {Griffiths, D. J.},
	year = {2013},
}

@article{Low2014,
	title = {Graphene plasmonics for terahertz to mid-infrared applications.},
	volume = {8},
	issn = {1936-086X 1936-0851},
	doi = {10.1021/nn406627u},
	language = {eng},
	number = {2},
	journal = {ACS nano},
	author = {Low, Tony and Avouris, Phaedon},
	month = feb,
	year = {2014},
	pmid = {24484181},
	pages = {1086--1101},
}

@Article{Grigorenko2012,
author={Grigorenko, A. N.
and Polini, M.
and Novoselov, K. S.},
title={Graphene plasmonics},
journal={Nature Photonics},
year={2012},
month={Nov},
day={01},
volume={6},
number={11},
pages={749-758},
issn={1749-4893},
doi={10.1038/nphoton.2012.262},
url={https://doi.org/10.1038/nphoton.2012.262}
}

@Article{Ju2011,
author={Ju, Long
and Geng, Baisong
and Horng, Jason
and Girit, Caglar
and Martin, Michael
and Hao, Zhao
and Bechtel, Hans A.
and Liang, Xiaogan
and Zettl, Alex
and Shen, Y. Ron
and Wang, Feng},
title={Graphene plasmonics for tunable terahertz metamaterials},
journal={Nature Nanotechnology},
year={2011},
month={Oct},
day={01},
volume={6},
number={10},
pages={630-634},
issn={1748-3395},
doi={10.1038/nnano.2011.146},
url={https://doi.org/10.1038/nnano.2011.146}
}

@article{Sohier2016,
  title = {Two-dimensional Fr\"ohlich interaction in transition-metal dichalcogenide monolayers: Theoretical modeling and first-principles calculations},
  author = {Sohier, Thibault and Calandra, Matteo and Mauri, Francesco},
  journal = {Phys. Rev. B},
  volume = {94},
  issue = {8},
  pages = {085415},
  numpages = {13},
  year = {2016},
  month = {Aug},
  publisher = {American Physical Society},
  doi = {10.1103/PhysRevB.94.085415},
  url = {https://link.aps.org/doi/10.1103/PhysRevB.94.085415}
}

@article{Sohier2017,
	title = {Breakdown of {Optical} {Phonons}’ {Splitting} in {Two}-{Dimensional} {Materials}},
	volume = {17},
	issn = {1530-6984},
	url = {https://doi.org/10.1021/acs.nanolett.7b01090},
	doi = {10.1021/acs.nanolett.7b01090},
	number = {6},
	journal = {Nano Letters},
	author = {Sohier, Thibault and Gibertini, Marco and Calandra, Matteo and Mauri, Francesco and Marzari, Nicola},
	month = jun,
	year = {2017},
	pages = {3758--3763},
}

@article{Ponce2023,
  title = {Accurate Prediction of Hall Mobilities in Two-Dimensional Materials through Gauge-Covariant Quadrupolar Contributions},
  author = {Ponc\'e, Samuel and Royo, Miquel and Gibertini, Marco and Marzari, Nicola and Stengel, Massimiliano},
  journal = {Phys. Rev. Lett.},
  volume = {130},
  issue = {16},
  pages = {166301},
  numpages = {6},
  year = {2023},
  month = {Apr},
  publisher = {American Physical Society},
  doi = {10.1103/PhysRevLett.130.166301},
  url = {https://link.aps.org/doi/10.1103/PhysRevLett.130.166301}
}

@article{Ponce2023b,
  title = {Long-range electrostatic contribution to electron-phonon couplings and mobilities of two-dimensional and bulk materials},
  author = {Ponc\'e, Samuel and Royo, Miquel and Stengel, Massimiliano and Marzari, Nicola and Gibertini, Marco},
  journal = {Phys. Rev. B},
  volume = {107},
  issue = {15},
  pages = {155424},
  numpages = {30},
  year = {2023},
  month = {Apr},
  publisher = {American Physical Society},
  doi = {10.1103/PhysRevB.107.155424},
  url = {https://link.aps.org/doi/10.1103/PhysRevB.107.155424}
}

@article{Royo2020b,
  title = {Using High Multipolar Orders to Reconstruct the Sound Velocity in Piezoelectrics from Lattice Dynamics},
  author = {Royo, Miquel and Hahn, Konstanze R. and Stengel, Massimiliano},
  journal = {Phys. Rev. Lett.},
  volume = {125},
  issue = {21},
  pages = {217602},
  numpages = {7},
  year = {2020},
  month = {Nov},
  publisher = {American Physical Society},
  doi = {10.1103/PhysRevLett.125.217602},
  url = {https://link.aps.org/doi/10.1103/PhysRevLett.125.217602}
}

@book{Martin2004, 
place={Cambridge}, 
title={Electronic Structure: Basic Theory and Practical Methods}, publisher={Cambridge University Press},
author={Martin,Richard M.},
year={2004}}

@article{Abinit2025,
	title = {Abinit 2025: {New} capabilities for the predictive modeling of solids and nanomaterials},
	volume = {163},
	issn = {0021-9606},
	url = {https://doi.org/10.1063/5.0288278},
	doi = {10.1063/5.0288278},
	number = {16},
	journal = {The Journal of Chemical Physics},
    author = {Verstraete, Matthieu J. and Abreu, Joao and Allemand, Guillaume E. and Amadon, Bernard and Antonius, Gabriel and Azizi, Maryam and Baguet, Lucas and Barat, Clémentine and Bastogne, Louis and Béjaud, Romuald and Beuken, Jean-Michel and Bieder, Jordan and Blanchet, Augustin and Bottin, Francois and Bouchet, Johann and others},
    month = oct,
	year = {2025},
	pages = {164126},
}

@article{Royo2019,
  title = {First-Principles Theory of Spatial Dispersion: Dynamical Quadrupoles and Flexoelectricity},
  author = {Royo, Miquel and Stengel, Massimiliano},
  journal = {Phys. Rev. X},
  volume = {9},
  issue = {2},
  pages = {021050},
  numpages = {22},
  year = {2019},
  month = {Jun},
  publisher = {American Physical Society},
  doi = {10.1103/PhysRevX.9.021050},
  url = {https://link.aps.org/doi/10.1103/PhysRevX.9.021050}
}

@article{Sohier2017b,
  title = {Density functional perturbation theory for gated two-dimensional heterostructures: Theoretical developments and application to flexural phonons in graphene},
  author = {Sohier, Thibault and Calandra, Matteo and Mauri, Francesco},
  journal = {Phys. Rev. B},
  volume = {96},
  issue = {7},
  pages = {075448},
  numpages = {21},
  year = {2017},
  month = {Aug},
  publisher = {American Physical Society},
  doi = {10.1103/PhysRevB.96.075448},
  url = {https://link.aps.org/doi/10.1103/PhysRevB.96.075448}
}

@article{Royo2022,
  title = {Lattice-mediated bulk flexoelectricity from first principles},
  author = {Royo, Miquel and Stengel, Massimiliano},
  journal = {Phys. Rev. B},
  volume = {105},
  issue = {6},
  pages = {064101},
  numpages = {17},
  year = {2022},
  month = {Feb},
  publisher = {American Physical Society},
  doi = {10.1103/PhysRevB.105.064101},
  url = {https://link.aps.org/doi/10.1103/PhysRevB.105.064101}
}

@article{Royo2021,
  title = {Exact Long-Range Dielectric Screening and Interatomic Force Constants in Quasi-Two-Dimensional Crystals},
  author = {Royo, Miquel and Stengel, Massimiliano},
  journal = {Phys. Rev. X},
  volume = {11},
  issue = {4},
  pages = {041027},
  numpages = {22},
  year = {2021},
  month = {Nov},
  publisher = {American Physical Society},
  doi = {10.1103/PhysRevX.11.041027},
  url = {https://link.aps.org/doi/10.1103/PhysRevX.11.041027}
}

@article{Gonze1997,
  title = {Dynamical matrices, Born effective charges, dielectric permittivity tensors, and interatomic force constants from density-functional perturbation theory},
  author = {Gonze, Xavier and Lee, Changyol},
  journal = {Phys. Rev. B},
  volume = {55},
  issue = {16},
  pages = {10355--10368},
  numpages = {0},
  year = {1997},
  month = {Apr},
  publisher = {American Physical Society},
  doi = {10.1103/PhysRevB.55.10355},
  url = {https://link.aps.org/doi/10.1103/PhysRevB.55.10355}
}

@article{Abinit2005,
author = {Xavier Gonze and Gian-Marco Rignanese and Matthieu Verstraete and Jean-Michel Beuken and 
        Yann Pouillon and Razvan Caracas and Francois Jollet and Marc Torrent and Gilles Zerah and Masayoshi Mikami
 and Philippe Ghosez and Marek Veithen and Jean-Yves Raty and Valerio Olevano and Fabien Bruneval and others},
         title = {A brief introduction to the ABINIT software package},
        journal = {Z. Kristallogr.},
        volume = {220},
        pages = {558},
        year = {2005},
        doi = {10.1524/zkri.220.5.558.65066},
}

@article{Li2024,
	title = {Observation of the nonanalytic behavior of optical phonons in monolayer hexagonal boron nitride},
	volume = {15},
	issn = {2041-1723},
	url = {https://doi.org/10.1038/s41467-024-46229-4},
	doi = {10.1038/s41467-024-46229-4},
	number = {1},
	journal = {Nature Communications},
	author = {Li, Jiade and Wang, Li and Wang, Yani and Tao, Zhiyu and Zhong, Weiliang and Su, Zhibin and Xue, Siwei and Miao, Guangyao and Wang, Weihua and Peng, Hailin and Guo, Jiandong and Zhu, Xuetao},
	month = mar,
	year = {2024},
	pages = {1938},
}

\end{document}